	\documentclass[preprint]{aastex63}

\received{2020 March 16}
\revised{2021 May 4}
\accepted{2021 May 19 }
\submitjournal{Astrophysical Journal} 

\shorttitle{Optical Properties of Interstellar Dust around Orion A}
\shortauthors{Uehara et al.}

\begin{document}
\title{Optical Properties of Interstellar Dust around the Orion A Molecular Cloud}

\correspondingauthor{Hayato Uehara}
\email{h.uehara@toho.ed.jp}
\author[0000-0002-6320-986X]{Hayato Uehara}
\affiliation{Department of Astronomy and Earth Sciences, Tokyo Gakugei University, 
4-1-1 Nukuikita, Koganei, Tokyo 184-8501, Japan}
\affiliation{TOHO Junior and Senior High School, 3-1-10 Naka, Kunitachi, Tokyo 186-0004, Japan }
\author[0000-0001-8058-8577]{Kazuhito Dobashi}
\affiliation{Department of Astronomy and Earth Sciences, Tokyo Gakugei University, 
4-1-1 Nukuikita, Koganei, Tokyo 184-8501, Japan}
\author[0000-0001-5898-5971]{Shingo Nishiura}
\affiliation{Department of Astronomy and Earth Sciences, Tokyo Gakugei University, 
4-1-1 Nukuikita, Koganei, Tokyo 184-8501, Japan}
\author[0000-0002-1054-3004]{Tomomi Shimoikura}
\affiliation{Faculty of Social Information Studies, Otsuma Women's University, 12 Sanban-cho, Chiyoda-ku, Tokyo 102-8357, Japan}
\author[0000-0001-6772-1628]{Takahiro Naoi}
\affiliation{Space Environment Laboratory, Applied Electromagnetic Research Institute, National Institute of Information and Communications Technology, 4-2-1 Nukuikita, Koganei, Tokyo 184-8795, Japan}



\begin{abstract}
We have studied optical properties of interstellar dust around the Orion A
molecular cloud to investigate the size distribution and the composition of dust grains.
Orion A is one of the most studied molecular clouds
in the solar vicinity ($d \simeq 400\ \rm{pc}$).
In this paper, we used optical and near-infrared photometric data.
The optical data were
obtained by $BVRI$ bands imaging observations.
The near-infrared data consisting of $JHK_{S}$ bands were taken from 2MASS point source catalog.
We produced some color excess maps around Orion A,
and measured their ratios such as $E(R-I)/E(B-V)$.
In order to investigate dust properties,
we compared the observed ratios with results of simulation
performed by Naoi T. et al. (2021)
who calculated the extinction in the optical to near-infrared wavelengths
based on a standard dust model;
they assumed a power-law grain-size distribution with an upper cutoff radius 
and assumed $graphite$ and $silicate$ as dominant components.
As a result, we found that the upper cutoff radius
around Orion A is $\simeq0.3$ $\micron$, 
and $silicate$ predominates compared with $graphite$
(with the fraction of $silicate$ grater than  $93$ \%).
In addition, we further derived the total-to-selective extinction ratio $R_{V}$ 
from the observed extinction of $A_{V}$ and the color excess $E(B-V)$,
and compared it with the model calculations.  
Dust properties
(i.e., the upper cutoff radius and the ratio of $graphite/silicate$)
derived from $R_{V}$ is almost consistent with those derived from the color excess ratios.
\end{abstract}
%
\keywords{Interstellar dust (836), 
		Interstellar dust extinction (837), 
		Diffuse interstellar clouds (380), 
		Extinction (505), 
		Reddening law (1377), 
		Optical observation (1169), 
		Near infrared astronomy (1093)}
\section{Introduction} \label{sec:intro}
The Orion molecular cloud complex is one of the best studied star-forming regions
in the solar neighborhood \citep[e.g.,][]{bally08}.
In this complex, there are two major giant molecular clouds called
Orion A and Orion B. 
Orion A includes Orion OB association 
which consists of four subgroups distributed around $b\simeq-19\degr$.
Ages of the subgroups are estimated to be 2 Myr to 12 Myr \citep{bla91,bro94}.
Additionally, there have been 10--20 supernova explosions 
during the last 12 Myr \citep{bally08}.
Some physical properties of the interstellar medium (ISM) are affected by
the UV-radiation from the OB associations 
and the shock waves from the supernovae \citep[e.g.,][]{bally87,wil05}.
In order to reveal physical properties of the clouds and interactions between stars and ISM, 
there have been a number of observations in various wavelengths.
Observations in the millimeter-wavelengths toward Orion A 
revealed the distributions and properties of the interstellar molecular gas 
leading to star formation \citep[e.g.,][]{nis15,nak19,ish19}, and
distributions and proper motions of young stars around Orion A have been widely studied
in the optical wavelengths \citep[e.g.,][]{Kounkel2018,Zari2019}.

Large scale extinction maps are useful to reveal the entire extents of
interstellar dust in Orion \citep[e.g.,][]{dob05,dob11}.
In recent years, a number of studies of interstellar dust around Orion A
have been carried out, utilizing accurate optical and near-infrared data provided by some
all sky survey projects.
For example, \cite{gre19} derived a three-dimensional map of dust, based on
Gaia \citep[e.g.,][]{gaia16,gaia18}, the Panoramic Survey Telescope and Rapid Response System 1
\citep[Pan-STARRS1; e.g.,][]{PS1}, and the Two Micron All Sky Survey \citep[2MASS;][]{str06}.
\citet{lom11} presented near-infrared extinction maps of large regions including
the Orion molecular clouds using the 2MASS Point Source Catalogue (2MASS-PSC).
They estimated the distance ($d=371\pm10$ pc), 
dust column density, and mass ($M=7.57 \times 10^{4} M_{\sun}$) of Orion A.
\citet{sch15} investigated the three-dimensional distribution of dust
toward the Orion molecular cloud complex 
on the basis of the Pan-STARRS1 photometry.
They also estimated the distance ($d=440\pm10$ pc) and
mass ($M=7.48 \times 10^{4} M_{\sun}$), 
and identified a ring-like structure in Orion A.
Three-dimensional dust distribution of the cloud complex
was also revealed by \citet{rez18}
using the Tycho-Gaia Astrometric Solution \citep[TGAS;][]{mic15,lin16},
near-infrared photometry from 2MASS-PSC, and mid-infrared photometry from 
Astronomical Wide-field Imaging System for Europe \citep[WISE;][]{wri10}.
They derived dust distribution in different distance ranges,
and showed three-dimentional structure of the Orion cloud complex which is distributed widely
in the range $300$ pc $\la d \la $ 550 pc.
\cite{gro18a,gro18b} attempted to reveal the three-dimensional shape
and orientation of Orion A
using distances of about 700 mid-infrared young stellar objects 
from the Gaia-DR2. \cite{gro18b} suggested that Orion A is not a fairly straight filamentary cloud,
but it is a cometary-like cloud oriented toward the Galactic plane
consisting of two distinct components:
the northern region (including the Orion nebula cluster) known as
a high-mass star-forming site is lying at $\sim 400$ pc,
while the southern region known as an inactive star-forming site
is inclined about $70\degr$ from the plane of the sky reaching at least $\sim 470$ pc.

General properties of interstellar dust have been long studied so far.
\cite{MRN} proposed that the observed interstellar extinction
of starlight may be caused
by a mixture of $graphite$ and $silicate$ particles having a simple power-law 
distribution of size: $n(r)\varpropto r^{-3.5}\ (0.005\ \micron \le r \le 0.25\ \micron)$
where $r$ is the grain radius.
\cite{dra84} re-investigated the extinction due to the mixture 
using the $graphite$ and $silicate$ dielectric functions.
They adopted the following power-law of dust size distribution: 
$dn_{i}=A_{i}n_{\rm{H}}r^{-3.5}dr$
where $n_{i}$ is the number density of grains of type $i$ with radius $\le r$
and $n_{\rm H}$ is the number density of H nuclei in both atoms and molecules.
They derived abundances of $graphite$ and $silicate$ per $n_{\rm H}$ in the Galaxy to be
$A_{graphite} = 10^{-25.16}$ 
and $A_{silicate} = 10^{-25.11}$.\ 
The ratio of these abundances is well known as a standard composition ratio of dust,
i.e., $graphite:silicate = 47 : 53$.
\citet{des90} proposed interpretation of both the interstellar extinction and the infrared emission,
using the dust model of minimal three components: 
Big Gains ($0.015\ \micron \le r \le 0.11\ \micron$),
Very Small Grains ($0.0012\ \micron \le r \le 0.015\ \micron$), and 
Polycyclic Aromatic Hydrocarbons (PAH ; $0.4\ {\rm nm} \le r \le 1.2\ {\rm nm}$) molecules.
Thereafter, \cite{zub04} tried to construct a unique interstellar dust model, 
but concluded that there is no unique interstellar dust model that simultaneously fits the observed extinction, 
diffuse IR emission, and abundance constraints.

There have been some studies to investigate growth of dust grains
in and around molecular clouds.
For example,
\citet{stp03} observed a quiescent dense filament in the Taurus molecular
cloud complex in the far-infrared wavelengths,
and found a change of dust properties inside and outside the filament,
and they pointed out the deficit of small grains in the dense filament
which is very likely to have been adsorped onto big grains.
\citet{kan03} investigated extinction and color excess of L1251, a dark cloud in Cepheus.
They found that the total-to-selective extinction ratio $R_{V}$ systematically and drastically changes
toward the densest part of the cloud, suggesting that the dust grains are
larger in the dense cloud interior.
\cite{mei18} also found that there is large variations of
reddening law in Orion. 

We have investigated the change of dust properties 
around the Taurus molecular complex in the optical and near-infrared wavelengths.
More recently, we compared the observed color excesses at different
bands with results of simulations performed by
\cite{nao19} who calculated the reddening law by dust based on the 
well-known dust models \citep{MRN,dra84} varying the size and components of dust,
and we concluded that the size of dust grains is
likely to be much larger inside the cloud than in the outside.
These results obtained for Taurus will be shown in a separate paper \citep{dob19}.s
In this paper, we investigate the dust properties (i.e., the size and
components) around Orion A by using the same methods as in Taurus.

To carry out this study, we obtained photometric data of stars
in seven bands in the optical and near-infrared ($BVRIJHK_{S}$),
and derived maps of various color excesses, extinction of $A_V$, 
and the total-to-selective extinction ratio $R_{V}$.
Details of the observations and the data reduction are described in Section \ref{sec:obs}.
In Section \ref{sec:res}, we describe how we produced 
the color excess maps and the extinction map.
In Section \ref{sec:DISCUSSION}, we compare the observational results with those of the model
calculations, and discuss the optical properties of dust.
Finally, we summarize the main conclusions in Section \ref{sec:CONCLUSION}.
\\
\section{Observations}	\label{sec:obs}
The optical data of Orion A were obtained in the period
from 2000 November to 2001 December 
using the 105 cm Schmidt Telescope
equipped with the 2kCCD camera \citep{ito01} at 
the Kiso Observatory belonging to the University of Tokyo. 
The camera has a CCD having 2048 $\times$ 2048 pixels.
The pixel resolution corresponds to 1$\farcs$5 on the sky,	
and the field of view of the camera is 51\farcm2 $\times$ 51\farcm2.
The camera was equipped with $B$, $V$, $R$,
and $I$ wide-band filters based on 
Johnson-Cousins photometric system \citep[e.g.,][]{bes90}.	
The limiting magnitude is 21, 22.5, 21, and 21 mag in 
S/N = 10 at a seeing size of 3\arcsec\ for the $B$, $V$, $R$, and $I$ band images, respectively.
We carried out $BVRI$ imaging observations of Orion A, and a region of
approximately $ 10 \degr \times 5 \degr $ centered at  $(l,b)=(211\degr,-18\degr)$ was covered.
We obtained an image at every 40\arcmin \ along the equatorial coordinates,
so that each image has an overlap of $\sim10\arcmin$ on the sky.
The exposure time was 200, 100, 100, and 50 sec in the $B$, $V$, $R$, and $I$ bands, respectively. 
Bias frames were obtained before and after the $BVRI$ imaging observations.
Dome flat frames were obtained in the beginning and at the end of each night.

For data reduction, we applied standard procedures including the subtraction
of bias current and flat fielding.
The defects in pixels (i.e., dead pixels) were masked.
We searched for point sources having intensities greater than 
the 5 $\sigma$ noise level from the sky background fluctuation, and regarded them as stars.
The typical seeing size (FWHM) of the detected stars was
{3\farcs6}, {3\farcs2}, {3\farcs4}, and {3\farcs1} in the
$B$, $V$, $R$, and $I$ bands, respectively. 
We applied aperture photometry and point spread function (PSF) photometry to the detected stars.
We established an image-by-image conversion relationship
between the PSF photometry and the aperture photometry,
and converted magnitudes of all stars measured
by the PSF photometry to those measured by the aperture photometry,
because the aperture photometry is more precise but only stars isolated from the others can be measured.
The pixel coordinates of the detected stars were converted into celestial coordinates,
comparing with their counterparts in the 2MASS-PSC.
A typical error in the coordinates is $\lesssim1\arcsec$.
We selected stars whose photometric error (1-$\sigma$ dispersion) is less than 0.05 mag in each band.
The mean photometric errors of the selected stars are
0.020, 0.019, 0.021, and 0.016 mag,
in the $B$, $V$, $R$, and $I$ bands, respectively.
Using transformation formulae obtained by \citet{ue19},
we converted the magnitudes of the detected stars
to those of the standard Johnson-Cousins photometric system.
To check the photometry, we further compared the converted
magnitudes with Pan-STARRS1 magnitudes whose transformation formulae
to the standard Johnson-Cousins system were obtained by \cite{ton12},
and found that they are consistent within the errors.
In addition, we searched for counterparts of the detected stars in the 2MASS-PSC
to obtain the $JHK_{S}$ data. 

\begin{figure}[ht!]
    \epsscale{0.80}
\plotone{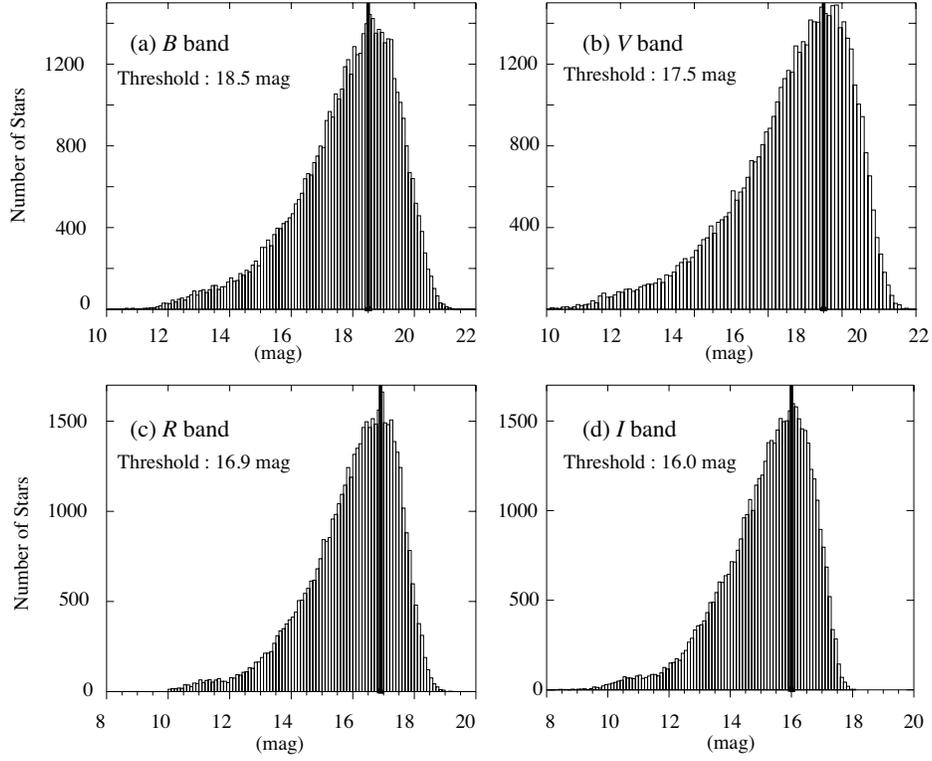}
\caption{The distributions of magnitudes in the $BVRI$ bands.
Bin size is 0.1 mag.
The solid lines indicate the threshold magnitudes.
}
 \label{fig:thre}
\end{figure}

Finally, we determined the threshold magnitudes giving complete detection in each band,
which will be used to produce maps of the color excess and the extinction.
Figure \ref{fig:thre} shows the distributions of magnitudes in the $BVRI$ bands
used to determine the threshold magnitudes.
The threshold magnitudes including the 2MASS $JHK_{S}$ bands are
summarized in Table \ref{tbl:thre}.

\section{Results}\label{sec:res}
	\subsection{Color Excess Maps}\label{sec:cem}
\begin{figure}[ht!]
  \plotone{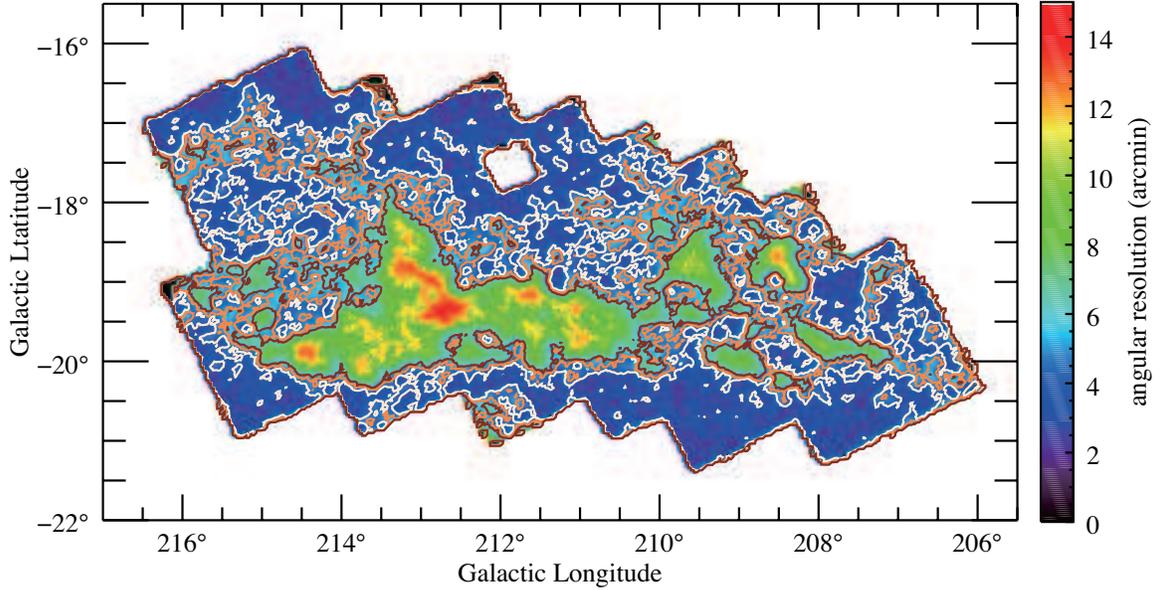}
\caption{Angular resolution map commonly used to produce the color excess
and the extinction maps in Figures \ref{fig:ebv} and \ref{fig:av}.
Contours are drawn at the angular resolution  8$\arcmin$, 10$\arcmin$, and 12$\arcmin$.
We analyzed only the region with the angular resolution higher than 10$\arcmin$.
We utilized this map to measure the mean star colors
such as $\overline{(B-V)}$
using stars within the angular resolution.
This map was also used when we counted stars to make an extinction map of $A_{V}$.
The white parts in this map denote the regions where there are no data.
}
\label{fig:beam}
\end{figure}

We made color excess maps of some combinations of colors,
utilizing a traditional method to measure the mean colors
of stars falling in a cell set along the celestial coordinates \citep[e.g., ][]{Lada1994}.
In the following, we explain how we made the color excess maps taking $E(B-V)$ as an example.
We first set grids at every 2$\arcmin$ along the Galactic coordinates,
and we measured the mean color $\overline{(B-V)}$ of the nearest $N$ stars to each grid.
The color excess $E(B-V)$ at each grid is defined as follows,
		\begin{equation}\label{eq:eq1}
			E(B-V) = \overline{(B-V)} - \overline{(B-V)}_{int}
		\end{equation}
where $\overline{(B-V)}_{int}$ is the mean value of the intrinsic star color without reddening.
In this work, we adopted the adaptive grid technique \citep{cam97}, 
and we measured the mean color of $N$ stars around each grid.
The angular resolution of the resulting map varies and 
is twice larger than the angle between the grid and the nearest $N$-th star.
We set a fixed number of $N=11$ to obtain an angular resolution of $\sim4\arcmin$--$20\arcmin$. 
The resolution varies also depending on the colors such as ($B-V$) and ($J-H$).
We therefore produced angular resolution maps of all of the combinations of colors for $N=11$, 
and found that the one for ($B-I$) 
gives the lowest resolution.
To standardize the angular resolutions of the various color excess maps,
we decided to use the resolution map of ($B-I$)
to measure all of the color excesses including $E(B-V)$ as a common resolution map. 
The common resolution map is shown in Figure \ref{fig:beam}.	
Contour levels in the figure indicates the angular resolution of
8$\arcmin$, 10$\arcmin$, and 12$\arcmin$.
We re-measured the color $\overline{(B-V)}$ at each grid
using the common angular resolution to reproduced the $E(B-V)$ map.

The mean value of the intrinsic star color $\overline{(B-V)}_{int}$ was obtained by the following process.
We utilized star data simulated at the website of Besan\c{c}on Galaxy Model
\footnote{Besan\c{c}on Galaxy Model Web site; https://model.obs-besancon.fr/}
\citep[BGM; e.g.,][]{BGM} covering the whole observed region,
and downloaded the star data without extinction to estimate the intrinsic star color $(B-V)_{int}$.
We measured the mean color of the BGM stars $\overline{(B-V)}_{BGM}$ at each grid in the same way
as we measured $\overline{(B-V)}$.
The value of $\overline{(B-V)}_{BGM}$ slightly fluctuates pixel by pixel, and
it may systematically change along with the Galactic coordinates.
In order to smooth the $\overline{(B-V)}_{BGM}$ map and to check the dependence on the coordinates,
we fitted $\overline{(B-V)}_{BGM}$ with the following function,
	\begin{equation}
			\overline{(B-V)}_{BGM} = C_{l} \cdot l + C_{b} \cdot b + C_{0}
		\label{eq:pq}
	\end{equation}
where $l$ and $b$ are the Galactic longitude and latitude, and 
$C_{l}$, $C_{b}$, and $C_{0}$ are the coefficients.
The best-fitted coefficients are summarized in Table \ref{tbl:pq}.
There is a slight dependence of the star color on $b$,
though the dependence on $l$ is rather negligible.
We regarded the fitted $\overline{(B-V)}_{BGM}$ map as the $\overline{(B-V)}_{int}$ map,
and obtained the final $E(B-V)$ map 
by subtracting the $\overline{(B-V)}_{int}$ map from the $\overline{(B-V)}$ map.

\begin{figure}[ht!]
    \epsscale{1.0}
  \plotone{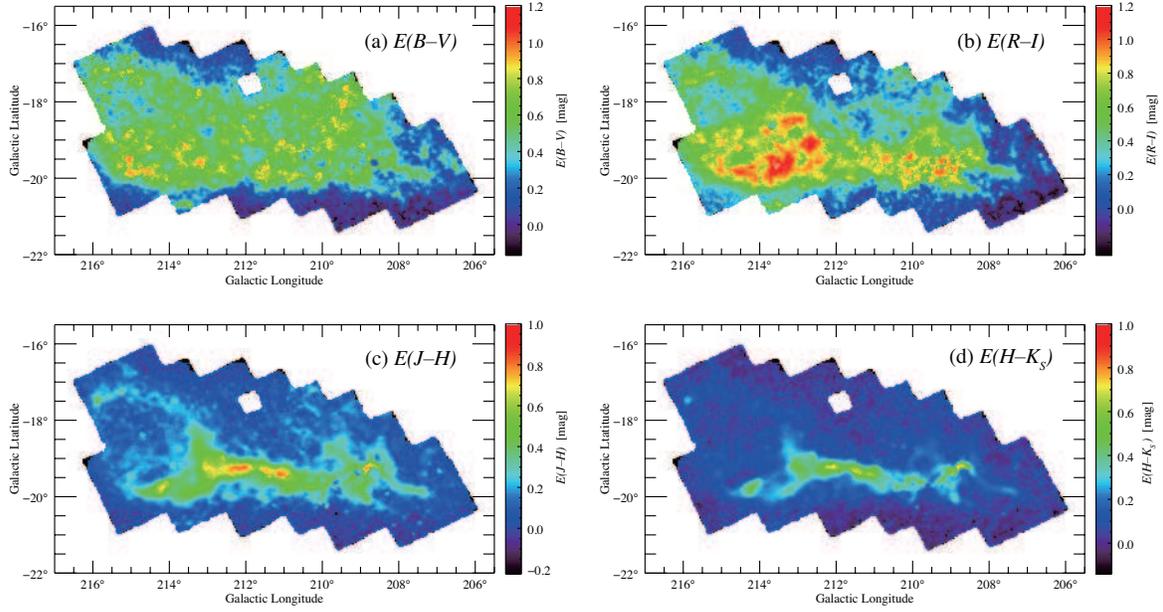}
\caption{The color excess maps of (a) $E(B-V)$, (b) $E(R-I)$, (c) $E(J-H)$,
and (d) $E(H-K_{S})$ obtained in this study.
}
\label{fig:ebv}
\end{figure}

In addition to the $E(B-V)$ map obtained in the above, 
we produced $E(R-I)$, $E(J-H)$, and $E(H-K_{\rm S})$ maps
in the same way.
Figure \ref{fig:ebv} shows the four color excess maps.	
We will use these maps in Section \ref{sec:Color Excess Ratio}.

When we prepared the simulated stars by the BGM, we neglected the diffuse interstellar
extinction along the line of sight, because it is difficult to assess precisely. This should
results in a systematic error in our color excess maps. If we take an average diffuse
interstellar extinction of 0.7 mag kpc$^{-1}$, this may cause an offset
of $\sim0.09$, $\sim0.08$, $\sim0.03$, and $\sim0.02$ mag
in the $E(B-V)$, $E(R-I)$, $E(J-H)$, and $E(H-K_{\rm S})$ maps
of Orion A located at $\sim400$ pc, respectively.

We performed Monte Carlo simulations to estimate errors due to counting uncertainty
in the color excess maps,
following the method described by \citet{dob11}.
We first composed probability distribution functions (PDFs)
of the intrinsic star colors in the observed region, such as $(B-V)$, using star data from the BGM.
We generated $N$ stars randomly whose colors follow the PDFs. Here, $N(\simeq11)$ is the number of
stars actually used to measure the mean star color at each grid. We repeated this procedure
1000 times for each grid, and measured the standard deviation of the mean colors, which we regard as
the noise levels of the color excess maps. Mean values of the resulting noise levels for the
$E(B-V)$, $E(R-I)$, $E(J-H)$, and $E(H-K_S)$ maps are
$\sigma_{E(B-V)}=0.104$,
$\sigma_{E(R-I)}=0.109$, 
$\sigma_{E(J-H)}=0.027$,
and
$\sigma_{E(H-K_S)}=0.025$ mag, respectively.

\subsection{Extinction Map of $A_{V}$}\label{sec:av}
We made an extinction map of $A_{V}$ from the photometric data of $V$ band by means of the star count method.
The grid spacing ($2\arcmin$) and angular resolution ($4\arcmin-20\arcmin$) of the map were set to
the same as those of the color excess maps.
Star density at each grid $D$ (i.e., the number of stars per unit solid angle), was estimated from
the number of stars within the resolution (Figure \ref{fig:beam}).

\begin{figure}[ht!]
  \plotone{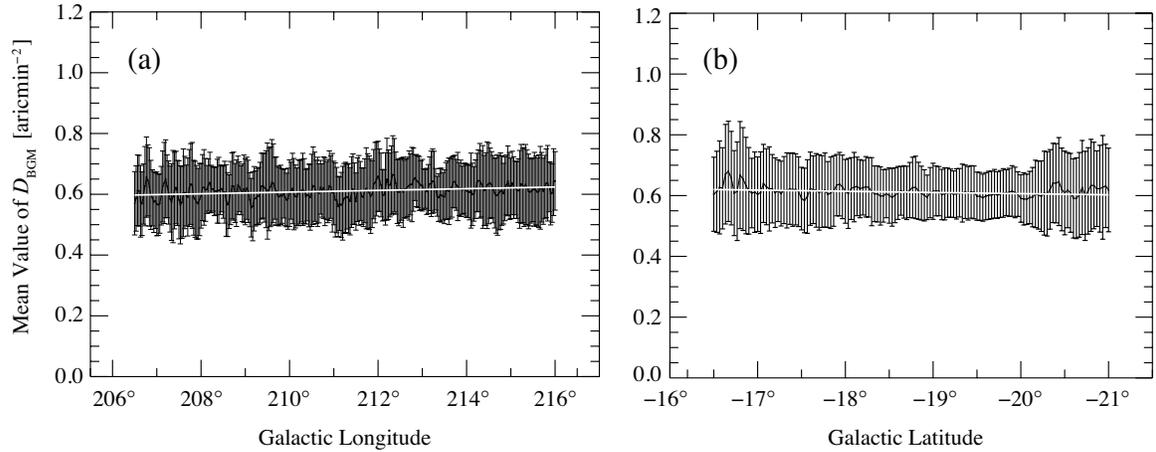}
\caption{Correlation between the mean value of $D_{BGM}$
and (a) the Galactic longitude and (b) the Galactic latitude.
The black line shows the mean value of $D_{BGM}$.
The error bars indicate the $1\sigma$ dispersion.
The white line indicates the fitted line by a linear function.
}
\label{fig:lbn}
\end{figure}

In order to obtain background stellar density $D_{back}$, 
we downloaded the photometric data of $V$ band from the BGM website without extinction.
We measured the star density of the BGM stars $D_{BGM}$ with the same methods for $D$,
and fitted $D_{BGM}$ as a function of the Galactic coordinates.
Background star density is often fitted with an exponential function of $b$ \citep[e.g.,][]{dob05}.
We inspected the dependence of $D_{BGM}$ on the Galactic coordinates ($l$ and $b$) around Orion A.
Black line in Figure \ref{fig:lbn} shows the dependence of the mean value of $D_{BGM}$ on $l$ and $b$.
The error bars indicate the $1\sigma$ dispersion, 
and the white line indicates the fitted line by a linear function.
As can be seen in Figure \ref{fig:lbn},
we found that change of $D_{BGM}$ can be well fitted by the linear functions
within the observed region ($205\degr \le l \le 216\degr$, $-22\degr \le b \le -16\degr$).
We therefore fitted $D_{BGM}$ with the following function,
	\begin{equation}
			D_{BGM} = C_{l}^{sc} \cdot l + C_{b}^{sc} \cdot b + C_{0}^{sc}
		\label{eq:nb}
	\end{equation}
where $C_{l}^{sc}$, $C_{b}^{sc}$, and $C_{0}^{sc}$ are the coefficients.
The best-fitted coefficients are
$C_{l}^{sc}=2.87\times10^{-3}$ mag $\deg^{-1}$,		
$C_{b}^{sc}=2.63\times10^{-4}$ mag $\deg^{-1}$, and	
$C_{0}^{sc}=0.010$ mag, respectively.					
We will regard the fitted $D_{BGM}$ as the background star density $D_{back}$.
 
\begin{figure}[ht!]
   \epsscale{0.70}
   \plotone{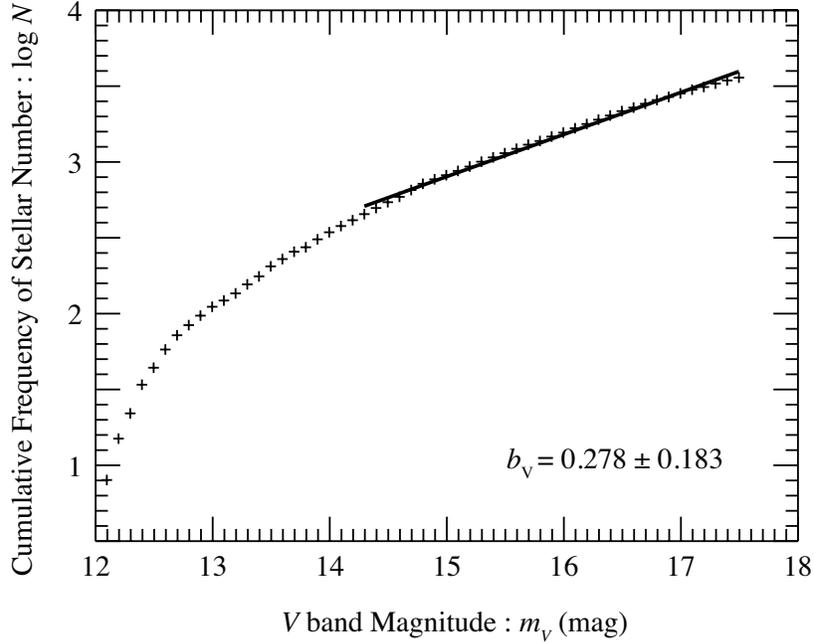}
\caption{Wolf diagram. The plus signs indicate the logarithmic cumulative frequency
measured in low extinction region ($206\degr \le l \le 211 \degr$
and $b \le {-20\fdg5}$).	
Bin size is 0.1 mag.
The black line indicates the best fitted line by Equation (\ref{eq:slo_wolf}). 
We obtained the slope $b_{V}$ to be 0.278 $\pm$ 0.183 mag$^{-1}$ 
in the range $14.3 \le m_{V} \le 17.5$ mag.
}
\label{fig:wolf}
\end{figure}

We constructed a cumulative frequency diagram of star number 
\citep[so-called ``Wolf diagram''; e.g.,][]{wolf} for the $V$ band magnitude
 as shown in Figure \ref{fig:wolf}.
In the figure, the bin size is 0.1 mag, and
the plus signs indicate the cumulative frequency obtained by our observations
which was measured in low extinction region ($206\degr \le l \le 211 \degr$	
and $b \le {-20\fdg5}$).													
In order to derive the slope of the Wolf diagram ($b_{V}$) necessary to calculate $A_{V}$,
we fitted the observed data as,
		\begin{equation}
		     \log{N}=b_{V} m_{V}+C_{1}
		\label{eq:slo_wolf}
		\end{equation}
where $m_{V}$ is the $V$ band magnitude and $C_{1}$ is a constant.
The best-fitted $b_{V}$ in Equation (\ref{eq:slo_wolf}) is obtained to be $0.278\pm 0.183$ mag$^{-1}$ 
in the range $14.3 \le m_{V} \le 17.5$ mag.

\begin{figure}
 \begin{center}
  \includegraphics[width=150mm]{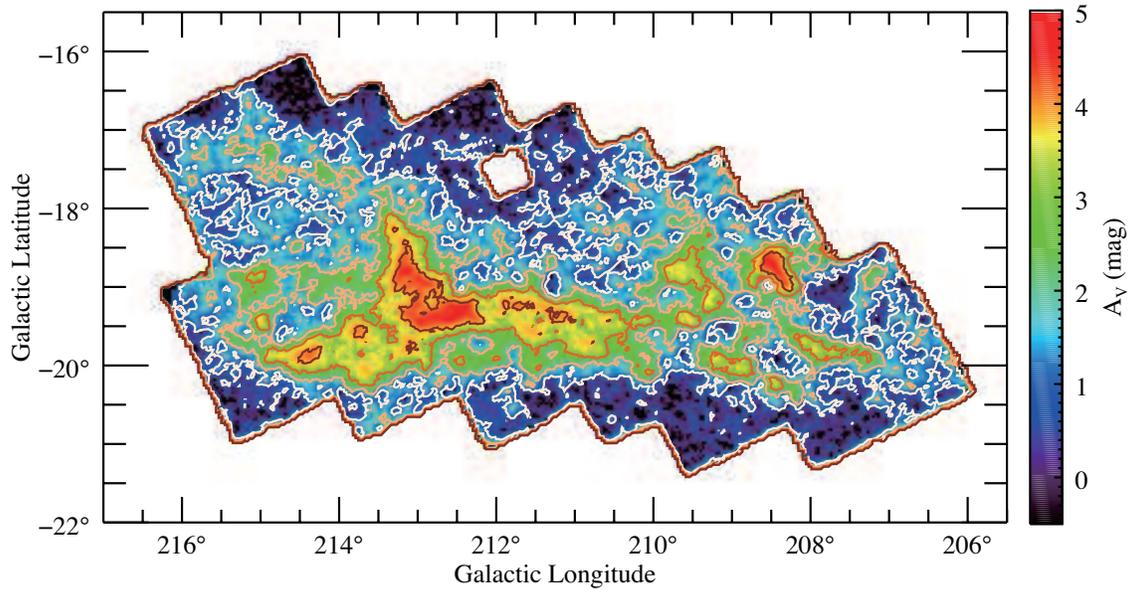}
 \end{center}
 \caption{Extinction map of $A_{V}$.
Contours are drawn from  $A_{V}=1$ mag to 4 mag with a step 1 mag.
The maximum extinction in the map amounts only to 
$A_{V}=4.94$ mag.The typical noise level of the $A_{V}$ map is 0.2 mag.
}
 \label{fig:av}
\end{figure}

Next, we calculated the extinction of $A_{V}$ for each grid as,
		\begin{equation}
			A_{V} = \frac{\rm1}{b_{V}}\log \left( \frac{D_{back}}{D}	\right).
		\end{equation}
We show the derived $A_{V}$ map in Figure \ref{fig:av}. 			

Finally, we also estimated the noise level of the $A_V$ map as,
 \begin{equation}
 \label{eq:av_err}
 \sigma_{A_{V}} = \frac{\rm log_{10}{\it e}}{b_{V}} \sqrt{\frac{1}{N}+\frac{1}{N_{back}}+\left(\frac{A_V}{\rm log_{10}{\it e}}\sigma_{b_{V}}\right)^2}
 \end{equation}
where
$N(\simeq11)$, $N_{back}$, and
$\sigma_{b_V}$$(\simeq 0.183$ mag$)$
are the number of stars actually used to count stars,
the number of the background stars in the same angular resolution estimated from the BGM, and
the standard error of  $b_{V}$, respectively. The noise level remains rather flat
with a mean value of
$ \sigma_{A_{V}}\simeq0.2$ mag,
because we used the adaptive grid method.
In addition to the above noise arising from the counting uncertainty,
there must be a systematic error in the $A_V$ map
caused by our neglecting the diffuse interstellar extinction when we derived the background
star density using the BGM. 
We estimate that this may cause an offset of $\sim0.1$ mag and $\sim0.4$ mag in our
$A_V$ map in the presence of the diffuse interstellar extinction of $0.7$ mag kpc$^{-1}$ and
$1.5$ mag kpc$^{-1}$, respectively.

The maximum value of the obtained map amounts only to
$A_{V}=4.94$ mag.
This is because the extinction can soon saturate
in the case of the optical wavelengths such as $V$ band,
and also because the star count method tends to underestimate
the extinction when there are a number of young stars forming
in the clouds. 
In this paper, we will analyze only the diffuse region
around Orion A where we are free from these errors.

\section{Discussion} \label{sec:DISCUSSION}
	\subsection{Color Excess Ratio}\label{sec:Color Excess Ratio}
The reddening and extinction are caused by the interstellar dust,
and they vary depending on the optical properties of the dust grains as well as on the wavelengths.
Utilizing this nature, we discuss the optical properties of dust grains from
ratios of different color excesses.
However, in regions where the star density is low due to high extinction,
it is difficult to measure precisely the color excesses and extinction
owning to the poor angular resolution that we can achieve.
From the analyses in the following,
we exclude the regions where the angular resolution is lower than 10\arcmin\ 
or the visual extinction of $A_{V}$ is grater than 2.0 mag,
and we will concentrate on the diffuse regions around Orion A.

\begin{figure}[ht]
 \begin{center}
  \includegraphics[width=80mm]{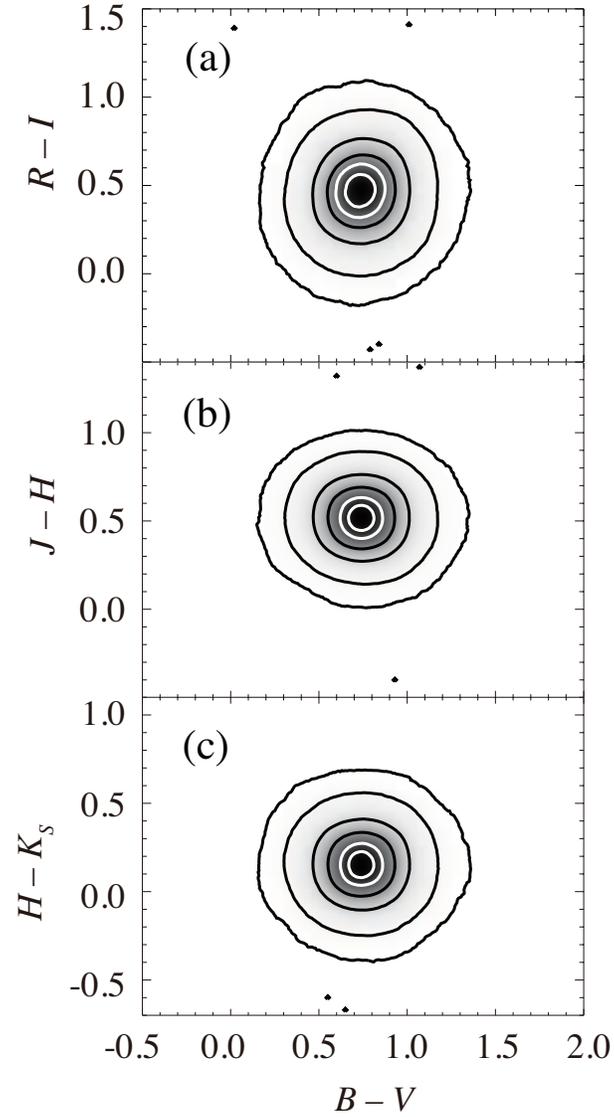}
 \end{center}
 \caption{Color vs. color diagrams of the intrinsic star colors 
simulated by the BGM. The common horizontal axis is $B-V$,
and the vertical axes are (a)$R-I$, (b)$J-H$, and (c)$H-K_{S}$. 
The data are shown by the plot densities and the contours
represent the levels where
99\% (the lowest contour), 95\%, 75\%, 50\%, 25\%, and
10\% (the highest contour) of the data points are included.
} 
 \label{fig:color}
\end{figure}

We will study the following three ratios of the color excesses:
 $E(R-I)/E(B-V)$, $E(J-H)/E(B-V)$, and $E(H-K_{S})/E(B-V)$.
The main reason for selecting these color excess ratios is that the variations of the intrinsic star colors
are almost symmetric, which is important to avoid artificial errors in measuring faint color excesses.
Figure \ref{fig:color} shows plot-densities of the intrinsic star colors		
which were derived from the BGM stars.
If the distributions are elliptic or asymmetrical in the figure,
it would affect the measurements of the faint color excesses.
Among various combinations of the color excesses,
we, therefore, selected those who show round and symmetric
shapes in color versus color diagrams.

\begin{figure}[ht]
\begin{center}
  \includegraphics[width=80mm]{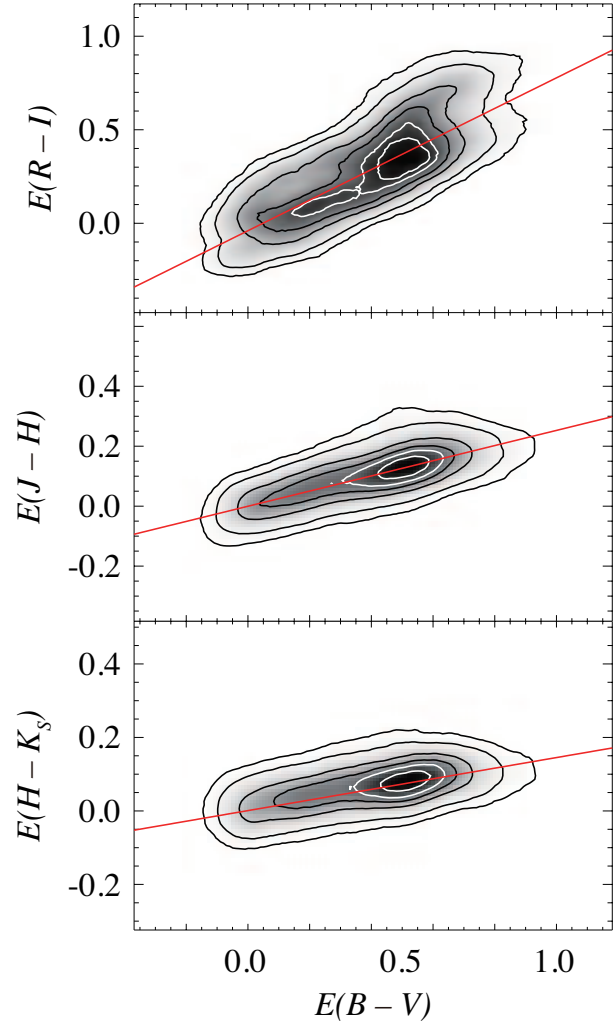}
 \end{center}
 \caption{Color excess vs. color excess observed around Orion A.
The common horizontal axis is $E(B-V)$, 
and the vertical axes are (a) $E(R-I)$, (b) $E(J-H)$, and (c) $E(H-K_{S})$.
Contours represent the plot densities of the data points
drawn at the same levels as in Figure \ref{fig:color}.
Solid lines are the linear relations in Equation (\ref{eq:kink})
obtained by the OLS method (Table \ref{tbl:cer}).
}
 \label{fig:ribv}
\end{figure}

Figure \ref{fig:ribv} shows the plot-density of the color excess ratios around Orion A.
In the figure, the common horizontal axis is $E(B-V)$,
and the vertical axes are (a) $E(R-I)$, (b) $E(J-H)$, and (c) $E(H-K_{S})$.
The linear relationships best fitting the color excesses are
shown by the red solid lines in the figure,
which can be expressed by the following formulae,
		\begin{equation}	
			E(\lambda_{1}-\lambda_{2}) =a_{0}+a_{1} \cdot E(B-V)  		\label{eq:kink}
		\end{equation}
where $E(\lambda_{1}-\lambda_{2})$ is the color excess between $\lambda_{1}$ band and $\lambda_{2}$ band, 
and $a_{1}$ represents the color excess ratio.
Although $a_{0}$ has to be zero in principle, it is actually not zero but has a slight value
($|a_{0}|=0.04$ mag at most),
which may be due to the systematic error in the determination
of the background star colors or the regression method used (see below).

Values of the obtained $a_0$ and $a_1$ are summarized in Table \ref{tbl:cer}.
The fit was performed using Ordinary Least Squares (OLS) method,
setting the weight $w$ using errors of the $E(\lambda_{1}-\lambda_{2})$ data (in the $Y$ axis)
as $w=1/\sigma_{E(\lambda_{1}-\lambda_{2})}^2$. Results of $a_0$ and $a_1$ do not change
significantly, even if we incorporate errors of the $E(B-V)$ data (the $X$ axis) setting the weight as
$w=1/[\sigma_{E(B-V)}^2+\sigma_{E(\lambda_{1}-\lambda_{2})}^2]$.

As shown by \cite{wan19}, fitting results can significantly change depending on the methods employed,
and we therefore checked how the values of $a_0$ and $a_1$ vary for different methods.
In addition to the OLS method,
we tried Standard Major Axis (SMA) method and Deming Method (DM). The former method
minimizes the total area of triangles formed by the observed ($X, Y$) data points and the regression line,
while the latter method minimizes the sum of the separation between the observed data points and
the regression line. Results are summarized in Table \ref{tbl:cer2}.
As can be seen in Tables \ref{tbl:cer} and \ref{tbl:cer2}, values of $a_1$ derived by the OLS
method are always lower than those derived by the other methods. The difference is 
large for $E(J-H)/E(B-V)$ and $E(H-K_S)/E(B-V)$, probably because that the slopes of the
$E(J-H)$ vs. $E(B-V)$ and $E(H-K_S)$ vs. $E(B-V)$ diagrams are rather small compared to the
dispersion of the data points (Figure \ref{fig:ribv}).
On the contrary, the slope of the $E(R-I)$ vs. $E(B-V)$ diagram is larger, and the
value of $a_1$  is better determined with a smaller difference among the fitting methods.

\cite{dob19} performed the same analyses around the Taurus molecular clouds, and
they found a ``knee'' or ``kink'' at $E(B-V)\simeq 0.3$ mag in most of the
relationships among the color excesses.
They reported that the color excess ratios are significantly different below and above the kink,
suggesting that the dust properties (such as the size) is different inside and outside the molecular cloud.
On the other hand, we could not find an obvious kink in Figure \ref{fig:ribv},
indicating that the dust properties remain rather constant around Orion A.

\cite{gro18a,gro18b} suggested that Orion A can be divided to the northern and southern regions
around at $(l,b)=(211.0\degr,-19.5\degr)$; the northern region has been known
to be a high-mass star-forming site and is lying at $\sim400$ pc,
while the southern region which has been known as an inactive star-forming site
is inclined about $70\degr$ from the plane of the sky reaching up to $\sim 470$ pc.
The two regions may have different properties of dust. 
We thus investigated the value of $a_1$ for the $E(R-I)/E(B-V)$ ratio in several smaller boxes
of $\sim 1$ deg$^2$ in the northern and southern regions. 
We found that $a_1$ tends to be
relatively higher
(typically $a_1 \simeq 0.8-1$)
in the northern region and lower ($a_1 \simeq 0.6-0.8$)
in the southern region, which may represent the difference of reddening law (and therefore the dust
properties) in the two regions. Actually, \citet{mei18} have studied infrared extinction in Orion A
with sensitive dataset at various wavelengths, and they found a significant variation of color
excess ratios in the cloud.

\subsection{Optical Properties of Interstellar Dust} \label{sec:rd}
\begin{figure}[ht]
 \begin{center}
  \includegraphics[width=80mm]{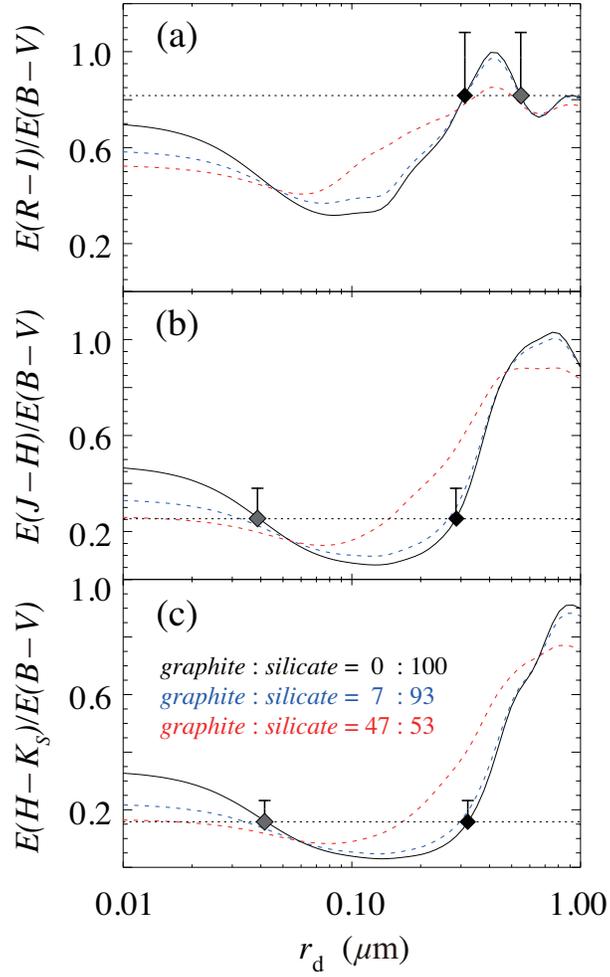}
 \end{center}
 \caption{
Relations of the color excess ratio $a_1$ 
and the upper cutoff radius of the size distribution $r_{\rm d}$ calculated
by \citet{nao19}. Panels (a), (b), and (c) show the ratios of
$E(R-I)/E(B-V)$, 
$E(J-H)/E(B-V)$, and 
$E(H-K_{S})/E(B-V)$, respectively.
Horizontal dotted lines denote $a_{1}$ in Table \ref{tbl:cer}.
Black solid and blue broken lines indicate the relations for 
the composition $graphite:silicate = 0 : 100$ and $7:93$.
Red broken lines indicate the relations for $graphite:silicate = 47 : 53$ 
(standard composition ratio, \citealt{dra84}).
Black diamonds indicate the solutions in $r_{\rm d}$ for the case of $graphite:silicate= 0 : 100$.
Error bars represent the variation of $a_1$ in Tables \ref{tbl:cer} and \ref{tbl:cer2}.
Ranges of the solution of $r_{\rm d}$ 
are summarized in Table \ref{tbl:rd}.
}
\label{fig:rd}
\end{figure}

The color excess ratio should reflect both the grain size and the composition of interstellar dust.
We inspect the grain size and the composition around Orion A based on the simulations performed
by \citet{nao19} who calculated the extinction based on the Mie scattering theory \citep{bh83}.
\citet{nao19} assumed spherical grains following the power-law of size-distribution suggested 
by \citet{MRN} with the upper cutoff $r_{\rm{d}}$ (i.e., the maximum grain radius of the size distribution), 
and adopted $graphite$ and $astronomical$ $silicate$ as dominant grain species \citep{dra84}. 
They calculated extinction for some standard photometric systems
(i.e., Johnson-Cousins, 2MASS, Gaia, and Pan-STARRS),
varying the composition ratio of $graphite:silicate$, 
which enable us to draw a plot of the color excess ratio
such as $E(R-I)/E(B-V)$ as a function of $r_{\rm{d}}$. 
An example is shown in Figure \ref{fig:rd}\,(a). 
By comparing our observational results with such plots,
we try to estimate $r_{\rm{d}}$ and the composition ratio of $graphite:silicate$ around Orion A.

In Figure \ref{fig:rd}, the horizontal dotted lines represent the observed color excess ratio
$a_1$ in Table \ref{tbl:cer} obtained by the OLS method, and the black solid, red broken,
and blue broken lines represent the
results derived from the calculation of \citet{nao19}.
The intersections (shown by the diamonds) of these lines give the solutions in $r_{\rm{d}}$.
As shown by the black diamonds in Figure \ref{fig:rd}\,(a),
there are often two solutions in $r_{\rm{d}}$ for a given set of the $a_1$ value
and the fraction of $silicate$.
One of the solutions gives the true $r_{\rm{d}}$, and the other is the false solution.
We expect that there must be a common solution in $r_{\rm{d}}$ for the three different color excess
ratios shown in panels (a)--(c) of the figure, if we find an appropriate fraction of $silicate$.

Within the variation of $a_1$ derived by the different regressions (Tables \ref{tbl:cer}
and \ref{tbl:cer2}), which is shown by the error bars in Figure \ref{fig:rd}, 
we investigated the ranges of the solution in $r_{\rm d}$ by changing the fraction of
$silicate$, and found that the three color excess ratios have a common solution in $r_{\rm d}$
only when the fraction of $silicate$ is greater than 93 \%. The ranges of $r_{\rm d}$ that we found
for the three color excess ratios within the range of the fraction of $silicate$ $93-100$ \% and
within the variation of $a_1$ are summarized in Table \ref{tbl:rd}.
As seen in the table, the common solution
can be found in the range $r_{\rm d}=0.312-0.328$ $\micron$.
We therefore conclude that the dust around Orion A has a fraction of $silicate$ greater than
93 \%  and $r_{\rm d}\simeq 0.3$ $\micron$.

We should note that this $r_{\rm d}$ value is the average over
Orion A, but the ambiguity of $a_1$ in Figure \ref{fig:ribv} is not small probably due to
the variation of the reddening law in smaller regions and also due to
unknown effects of the creation of mantle which is not considered in our analyses.
However, we suggest that
$r_{\rm d}\simeq 0.3$ $\micron$ or a little higher value
would be plausible as the average dust size around Orion A,
because it is consistent with theoretical calculations and some observations (see below).
It is also interesting to note that the $r_{\rm d}$ value is significantly larger than that found
around the Taurus molecular cloud \citep[$r_{\rm d}\simeq 0.17$ $\micron$,][]{dob19}, suggesting that
dust grains in Orion A are more grown and evolved probably through coagulation by collisions. 

The high fraction of $silicate$ of this dust size ($r_{\rm d}\simeq0.3$ $\micron$)
can be consistent
with theoretical calculations by \cite{we01}. They suggested that dust is much richer in $silicate$ than in $graphite$
for the grain size $r_{\rm d} \simeq 0.1 - 0.3$ $\micron$.
The fraction of $silicate$ has a peak around the grain size $0.2-0.3$ $\micron$,
and rapidly decreases for larger size, being almost zero around $\sim 0.4$ $\micron$.
They also showed that the fraction of $silicate$ changes depending on the total-to-selective
extinction ratio $R_V$. 
Looking at the plots of their calculations shown in their Figures 2 -- 6 by eyes, 
the values of $R_V=3.1$, $4.0$, and $5.5$ yield the fraction of $silicate$ of
$\sim30$ \%,  $\sim 60$ \% and $\sim90$ \%, respectively,
at the grain size $r_{\rm d}=0.3$ $\micron$.
We derived $R_V$ around Orion A to be roughly $\gtrsim 4$ (Section $\ref{sec:Rv}$).
The corresponding fraction of $silicate$ inferred from their calculations ($\gtrsim 60$ \%) is therefore
lower than what we found in this study ($\gtrsim 93$ \%), but they are qualitatively consistent in
a sense that $silicate$ is the dominant component rather than $graphite$.

Our results are also consistent with other previous studies
that most of big grains (with $r_{\rm{d}} \gtrsim 0.1\ \micron$) are composed of $silicate$
\citep[e.g.,][]{des90,zub04}.
In addition, it is noteworthy that the emission of amorphous (possibly crystalline)
$silicate$ grains is widely observed over the Orion H{\scriptsize{II}} region
and extended region around the O9.5Vpe star $\theta^{2}$ Ori A \citep{ces00}.

\subsection{Total-to-Selective Extinction Ratio $R_{V}$}\label{sec:Rv}
\begin{figure}[ht]
 \begin{center}
  \includegraphics[width=90mm]{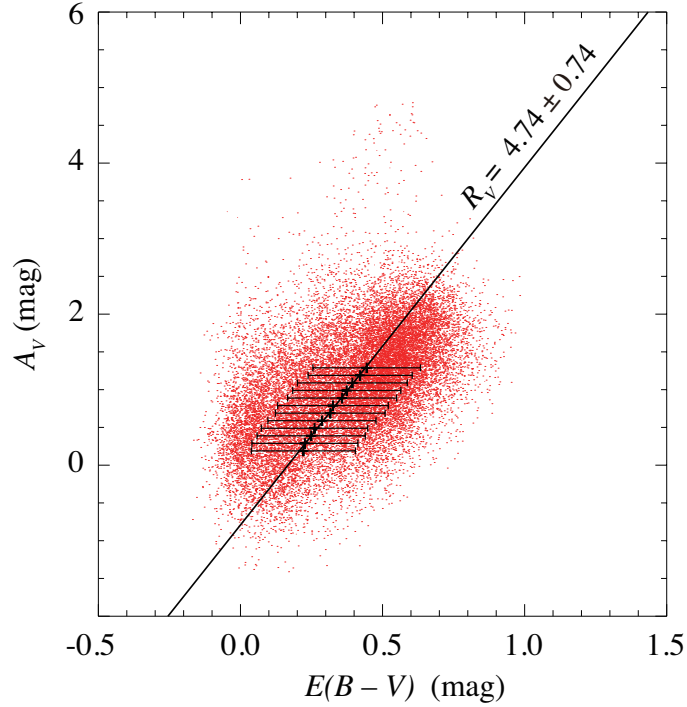}
 \end{center}
 \caption{Correlation between $A_{V}$ and $E(B-V)$.
The red dots indicate the data of the entire observed region.
Plus signs indicate mean values of $E(B-V)$ in the bins of $\Delta A_{V}=0.1$ mag, 
and error bars indicate the 1$\sigma$ dispersion.
The linear relation given by Equation (\ref{eq:rv}) best fitting the plus
signs is shown by the solid line.
}
  \label{fig:rva}
\end{figure}

Extinction of $A_{V}$ is proportional to the color excesses including $E(B-V)$.
It is well-known that the total-to-selective extinction ratio $R_{{V}}$ is
an indicator both of the grain size and the composition of the interstellar dust \citep[e.g.,][]{MRN}.
$R_{{V}}$ is defined as
\begin{equation}\label{eq:rv0}
	A_{V} = R_{V}\cdot E(B-V) ~~.
\end{equation}
Figure \ref{fig:rva} shows the pixel-to-pixel relationship between $E(B-V)$ and $A_{V}$
in the observed region.
In the figure, the red dots indicate the data of the entire observed region.
As seen in the figure, the dispersion of the data point is large.
We tried to measure $R_{{V}}$ by fitting the red dots with Equation (\ref{eq:rv0}) and
Equation (\ref{eq:rv}) in the following using some
different types of regressions, but it was difficult to obtain reliable results because of the low 
correlation coefficients ($\simeq 0.51$ at most).
We therefore divided the $A_{V}$ range into 0.1 mag bins in the range from 0 to 1.2 mag,
and calculated the mean values of $E(B-V)$ in each bin.
In Figure \ref{fig:rva}, the mean values of $E(B-V)$ are shown by plus signs.
The error bars indicate 1$\sigma$ dispersion.
We fitted the plus signs using, 	
\begin{equation}\label{eq:rv}
	A_{V} = R_{V}\cdot E(B-V)+C_{\rm zero}
\end{equation}
where $C_{\rm zero}$ is a constant to adjust an offset in our $A_V$ map.
The best-fitted $R_{V}$ we obtained is $4.74\pm0.74$, and we show the results
in Figure \ref{fig:rva} and Table \ref{tbl:rv}.
The value of $R_{V}$ is about 50\% larger than the average value $R_{V}=3.1$	
in the solar neighborhood \citep[e.g.,][]{fm07}.
The fit was made with the OLS method. We also tried other regressions (SMA and DM)
as we performed in Section \ref{sec:Color Excess Ratio}, and found consistent results
with those obtained by the OLS method.

The fitted value of $C_{zero}$ is $-0.79$ mag, which is rather large.
We consider that this should be mainly caused by the same effect
that we mentioned in Section \ref{sec:rd} for the possible errors
in the color excess ratios (i.e., $a_1$); we fitted all of the data points
in the large observed region at once, while there are certain
variations in smaller local regions. In addition, we neglected
the diffuse interstellar extinction when we determine the background
star density to derived the $A_V$ map (see Section \ref{sec:av}),
which is also a cause of $C_{zero}\ne 0$.

\begin{figure}[ht]
 \begin{center}
  \includegraphics[width=80mm]{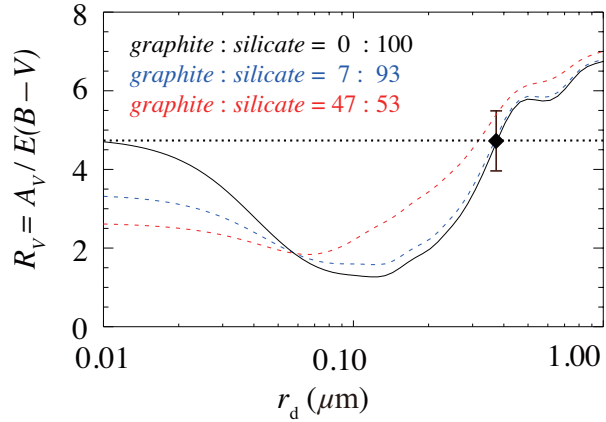}
 \end{center}
 \caption{Comparison of $R_{V}$ with $r_{\rm{d}}$ based on the model calculations.
$R_{V}$ found in Figure \ref{fig:rva} ($R_V=4.74$)
 is drawn by the horizontal dotted line.
Black solid and blue broken lines indicate the relation for the composition ratios $graphite:silicate= 0 : 100$
and  $7 : 93$, respectively.
Red broken line is for the standard composition ratio $graphite:silicate= 47 : 53$ (\citealt{dra84}).
Black diamonds denotes the solutions in $r_{\rm d}$ for the case of $graphite:silicate= 0 : 100$.
Ranges of the solution of $r_{\rm d}$ 
are summarized in Table \ref{tbl:rv}.}
  \label{fig:rvb}
\end{figure}

Figure \ref{fig:rvb} shows the model-based diagram of the relationship
between $R_{V}$ and $r_{\rm{d}}$ derived from the results of calculations by \citet{nao19}.
We tried to derive the $r_{\rm d}$ value based on this diagram.
As we did in Section \ref{sec:rd}, intersections of the model curves and the observed
$R_V$ value give the solutions in $r_{\rm d}$. Within the range 
of the fraction of $silicate$ 93 -- 100 \%, the solution varies in the range
$r_d=0.326$ -- $0.431$ $\micron$, which is also summarized in Table \ref{tbl:rv}.
Though the determination of $R_V$ apparently suffers from a large uncertainty
because of the large dispersion of the data points, it is noteworthy that this range of
$r_d$ is consistent with
those obtained from the analyses of the color excess ratios in Section \ref{sec:rd}.
Note also that a small error in determining $R_V$ may not change the range
of $r_{\rm d}$ significantly, because the model curve of $R_V$ for the fraction of
$silicate\ge 93$ \%
is steep around $r_{\rm d}=0.3$ $\micron$ (see Figure \ref{fig:rvb}).
To conclude, we therefore suggest that the average dust grains around Orion A are large in size
(with $r_{{\rm d}}\simeq 0.3$ $\micron$) and they consist mainly of
$astronomical$ $silicates$ rather than $graphite$ with a fraction of $silicate$ $\gtrsim 93$ \%.

\section{Conclusions}\label{sec:CONCLUSION}
In order to investigate the dust properties around the Orion A molecular cloud,
we produced the color excess maps and an extinction map of $A_{V}$,
and derived the ratios of color excesses such as $E(R-I)/E(B-V)$ and
the total-to-selective extinction ratio $R_{V}$.
We compared these ratios 
with results of model calculations performed by \cite{nao19} to infer the size and components
of dust grains around Orion A.
The model adopted by \cite{nao19} assumes a power-law size-distribution of dust
with the upper cutoff radius $r_{\rm d}$. 
Main findings of this study can be summarized as follows:

\begin{enumerate}
\item{Comparison of the observed ratios of color excesses
such as $E(R-I)/E(B-V)$ with the prediction by the model infers that $r_{\rm d}$ is $\simeq 0.3$ $\micron$
for dust around Orion A. This value of $r_{\rm d}$ is significantly larger than that observed in
Taurus, suggesting that dust grains are grown and more evolved than in Taurus
probably through coagulation by collisions.}	

\item{Comparison with the model calculations also infers that the observed ratios of the color
excesses can be fitted by the model for dust consisting mainly of $silicate$ rather than $grapthie$
with a fraction of $silicate$ $\gtrsim 93$ \%. The fraction of $silicate$ is very different from
the standard composition ratio of dust \citep[$graphite:silicate=47 : 53$,][]{dra84}.
We suggest that the dominant component of dust around Orion A is $silicate$, which
is consistent with results of earlier observations.}

\item{The $R_{V}$ value derived directly from the observed $A_V$ and $E(B-V)$
around Orion A is
$4.74 \pm 0.74$.
Though dispersion of the $A_V$ vs. $E(B-V)$
diagram is large, comparison of this $R_V$ value with the model
calculations infers the dust size
consistent with 
that inferred} from the color excess ratios.			
\end{enumerate}
	
\acknowledgments
We thanks the kind supports of the staffs at Kiso Observatory during observations.
HU is grateful to Tokyo Gakugei University and TOHO junior and senior high school for
support and hospitality during his sabbatical leave.
This work was financially supported by JSPS KAKENHI Grant Numbers 17H02863, 17H01118, and 19H05070.	




%
\begin{deluxetable}{cccc}
\tablecaption{Threshold Magnitudes of Each Band		\label{tbl:thre}}
\tablewidth{0pt}
\tablehead{
\colhead{Band} & \colhead{Threshold (mag)} & \colhead{Band} & \colhead{Threshold (mag)}
}
\startdata
  $\it{B}$	&	18.5	&	$\it{V}$	&	17.5	\\
  $\it{R}$	&	16.9	&	$\it{I}$	&	16.0	\\  
  $\it{J}$	&	16.7	&	$\it{H}$	&	15.8	\\
  $\it{K}_{\rm{S}}$	&	15.3	\\	[2pt] 
\enddata
\end{deluxetable}
%
\begin{deluxetable}{cccc}
\tablecaption{Coefficients of Equation (\ref{eq:pq})			\label{tbl:pq}}
\tablewidth{0pt}
\tablehead{
\colhead{Color} & \colhead{$C_{l}$} & \colhead{$C_{b}$} & \colhead{$C_{0}$} \\
\nocolhead{}	& \colhead{($\times 10^{-4}$ mag $\deg^{-1}$)}
				& \colhead{($\times 10^{-3}$ mag $\deg^{-1}$)}	    & \colhead{(mag)}	
}
\decimals
\startdata
  $B-V$				&	$-0.812$	& $ -3.0162 $	&	0.7026	\\
  $R-I$				&	$-7.538$	& $  0.6889	$   &	0.6101	\\  	
  $J-H$				&	$ 1.425$	& $ -1.7608 $   &	0.4536	\\
  $H-K_{\rm{S}}$	&	$ 4.425$	& $  0.7005 $	&	0.0755	\\
\enddata
\end{deluxetable}

\begin{deluxetable}{ccc}
\tablecaption{Color Excess Ratios Obtained by the Ordinary Least Squares Method \label{tbl:cer}}
\tablewidth{0pt}
\tablehead
{
\nocolhead{}				  	&	Slope						 & Intercept\\	
\colhead{Color Excess Ratio}  &   $a_{1} \pm \sigma_{a_{1}}$ 		 &	$a_{0} \pm \sigma_{a_{0}}$
}
\startdata
  $E(R-I)/E(B-V)$			&	$0.817 \pm 0.0052$	&	$-0.040  \pm 0.0020$	\\  	
  $E(J-H)/E(B-V)$			&	$0.253 \pm 0.0057$	&	$-0.0003 \pm 0.0025$	\\  	
  $E(H-K_{\rm S})/E(B-V)$	&	$0.144 \pm 0.0045$	&	$ 0.0009 \pm 0.0019$	\\  	
 \enddata
\tablecomments{The values of $a_{0}$ and $a_{1}$ are obtained by the Ordinary Least Squares (OLS) method,
and they correspond to those in Figure \ref{fig:ribv}.
}
\end{deluxetable}

\begin{deluxetable}{cccc}
\tablecaption{Color Excess Ratios obtained by Other Regressions \label{tbl:cer2}}
\tablewidth{0pt}
\tablehead
{
\nocolhead{}				 &  Regression 	&	Slope						 & Intercept\\	
\colhead{Color Excess Ratio} & Method 		&$a_{1} \pm \sigma_{a_{1}}$ 		 &	$a_{0} \pm \sigma_{a_{0}}$
}
\startdata
   $E(R-I)/E(B-V)$			& SMA		&	$1.039 \pm 0.0039$	&	$ 0.048 \pm 0.0017$	\\
  							& DM		&	$1.081 \pm 0.0055$	&	$-0.120 \pm 0.0022$	\\ \cline{1-4}
  $E(J-H)/E(B-V)$			& SMA		&	$0.392 \pm 0.0208$	&	$-0.044 \pm 0.0219$	\\
 							& DM		&	$0.267 \pm 0.0013$	&	$-0.006 \pm 0.0006$	\\ \cline{1-4}
  $E(H-K_{\rm S})/E(B-V)$	& SMA		&	$0.218 \pm 0.0048$	&	$-0.017 \pm 0.0084$	\\
  							& DM		&	$0.155 \pm 0.0014$	&	$-0.002 \pm 0.0006$	\\
\enddata
\tablecomments{Values of $a_{0}$ and $a_{1}$ are obtained by the Standard Major Axis (SMA) regression and
Deming Method (DM).
}
\end{deluxetable}
%

\begin{deluxetable}{ccccc}
\tablecaption{Upper Cutoff Radius $r_{\rm d}$ of Dust Size Distribution	\label{tbl:rd}}
\tablewidth{0pt}
\tablehead
{
\colhead{Combinations  }	& & \multicolumn{2}{c}{Range of Upper Cutoff Radius}	\\	
\cline{3-4} 
\colhead {of } 					&  &Lower $r_{\rm d}$	& Higher $r_{\rm d}$\\ 
\colhead{Color Excess }		& &	(\micron)			&  (\micron)
}
\startdata
  $E(R-I)/E(B-V)$			& &	  0.312 -- 0.402   &   (0.415 -- 0.549)	\\  	
  $E(J-H)/E(B-V)$			& &  (0.023 -- 0.039)	&	 0.263 -- 0.328	\\
  $E(H-K_{\rm S})/E(B-V)$	& &  (0.011 -- 0.045)	&	 0.282 -- 0.354	\\
\enddata
\tablecomments{Ranges of $r_{\rm d}$ which vary depending on the fraction of
$silicate$ (for the range $93\% - 100\%$) 
and the values of $a_1$ (Tables \ref{tbl:cer} and \ref{tbl:cer2})
are summarized. For a given set of the fraction of $silicate$ and $a_1$, there are two solutions in $r_{\rm d}$
(Figure \ref{fig:rd}), as they are indicated as ``Lower $r_{\rm d}$"  and ``Higher $r_{\rm d}$" in the table.
Ranges of $r_{\rm d}$ in the parentheses are false solutions (see text).
}
\end{deluxetable}

\begin{deluxetable}{cccc}
\tablecaption{Upper Cutoff Radius $r_{\rm d}$ and $R_{V}$	\label{tbl:rv}}
\tablehead
{
\colhead{Regression}	& \colhead{$R_{V} \pm \sigma R_{V}$}   & \colhead{Upper cutoff radius}	\\
\colhead{Method}	& \colhead{} & $r_{\rm d}$ $(\micron)$	
}
\startdata
 			&	$4.74\pm 0.78	$	&  0.326 -- 0.431	\\	
\enddata
\tablecomments{$R_V$ and $r_{\rm d}$ derived by the analyses shown
in Figures \ref{fig:rva} and \ref{fig:rvb}
for the range of the fraction of $silicate$ $93\% - 100\%$.
}
\end{deluxetable}

%
%
%
%
%

%

%

%
%

\end{document}